\documentclass[10pt]{iopart}
\usepackage{iopams}  
\usepackage{graphicx}
\begin{document}

\title[Local structure in Ag$_{\it3}{\textit\lbrack}$Co(CN)$_{\it6}{\textit\rbrack}$]{Local structure in Ag$_3$[Co(CN)$_6$]: Colossal thermal expansion, rigid unit modes and argentophilic interactions}

\author{Michael J Conterio$^1$, Andrew L Goodwin$^{1}$, Matthew G Tucker$^2$, David A Keen$^{2,3}$, Martin T Dove$^1$, Lars Peters$^4$ and John S O Evans$^4$}

\address{$^1$ Department of Earth Sciences, University of Cambridge, Downing Street, Cambridge CB2 3EQ, U. K.}
\address{$^2$ ISIS Facility, Rutherford Appleton Laboratory, Harwell Science and 
Innovation Campus, Didcot, Oxfordshire OX11 0QX, U.K.}
\address{$^3$ Department of Physics, Oxford University, Clarendon Laboratory, 
Parks Road, Oxford OX1 3PU, U.K.}
\address{$^4$ Department of Chemistry, University of Durham, University Science 
Laboratories, South Road, Durham DH1 3LE, U.K.}
\ead{alg44@cam.ac.uk}
\begin{abstract}
Local structure in the colossal thermal expansion material Ag$_3$[Co(CN)$_6$] is studied here using a combination of neutron total scattering and reverse Monte Carlo (RMC) analysis. We show that the large thermal variations in cell dimensions occur with minimal distortion of the [Co(CN)$_6$] coordination polyhedra, but involve significant flexing of the Co--CN--Ag--NC--Co linkages. We find real-space evidence in our RMC configurations for the importance of low-energy rigid unit modes (RUMs), particularly at temperatures below 150\,K. Using a reciprocal-space analysis we present the phonon density of states at 300\,K and show that the lowest-frequency region is dominated by RUMs and related modes. We also show that thermal variation in the energies of Ag$\ldots$Ag interactions is evident in both the Ag partial pair distribution function and in the Ag partial phonon density of states. These findings are discussed in relation to the thermodynamic properties of the material.
\end{abstract}

\pacs{PACS}
\submitto{\JPCM}
\section{Introduction}

Flexibility plays a key role in determining the physical properties of framework materials. In many oxide-containing frameworks, for example, the underconstraint of linear cation--oxygen--cation (M--O--M) linkages means that the M--O--M angle is especially sensitive to changes in temperature, pressure and composition. A direct consequence of this underconstraint is the rich series of polyhedral-tilting transitions known to pervade the broad and technologically-important families of perovskite materials \cite{Mitchell_2002} and silicate frameworks \cite{Hammonds_1996,Dove_1997}. A similar example is the well-known negative thermal expansion (NTE) material ZrW$_2$O$_8$, where the same local flexibility of M--O--M linkages is implicated in a number of unusual physical properties: NTE itself \cite{Mary_1996,Pryde_1996,Tucker_2005}, elastic constant softening under pressure \cite{Pantea_2006}, and pressure-induced amorphisation \cite{Keen_2007}.

It is becoming increasingly evident that the two-atom metal--cyanide--metal (M--CN--M) linkages found in many transition metal cyanides \cite{Sharpe_1976} can produce a similar, and often more extreme, array of unusual physical behaviour. Because the cyanide molecule is able to translate as an independent unit, not only is the linkage especially flexible but much of the correlation between displacement patterns around neighbouring M centres is lost \cite{Goodwin_2006}. What this means is that properties such as NTE appear to be inherently more common amongst cyanides than amongst oxide-containing frameworks, and also the magnitude of these effects can be much larger \cite{Hibble_2002,Margadonna_2004,Goodwin_2005,Goodwin_2005b,Pretsch_2006}. For example, isotropic NTE behaviour in the Zn$_x$Cd$_{1-x}$(CN)$_2$ family is more than twice as strong, and in the case of single-network Cd(CN)$_2$ more than three times as strong, as that in ZrW$_2$O$_8$ \cite{Goodwin_2005,Phillips_2008}.

The concept of increasing framework flexibility---and hence the likelihood and extent of unusual physical behaviour---by employing linear structural motifs of increasing length has led us recently to discover colossal positive and negative thermal expansion in Ag$_3$[Co(CN)$_6$] \cite{Goodwin_2008}. The covalent framework structure of this material, discussed in more detail below, is assembled from Co--CN--Ag--NC--Co linkages, so that the transition metal centres are now connected by linear chains containing five atoms. The framework lattice produced is so flexible that it can adopt a wide variety of geometries with essentially no difference in lattice enthalpy. This serves to amplify the thermodynamic role of low-energy bonding interactions, in this case the dispersion-like ``argentophilic'' interactions between neighbouring Ag$\ldots$Ag pairs. The net effect is that the lattice expands in some directions at a rate typical of those observed in van-der-Waals solids (\emph{e.g.}\ Xe \cite{Sears_1962}) and this expansion is coupled via flexing of the framework lattice to an equally strong NTE effect along a perpendicular direction: the coefficients of thermal expansion, $\alpha$, were found to be roughly linear and equal to $\pm130$\,MK$^{-1}$ (1\,MK$^{-1}=10^{-6}$\,K$^{-1}$) over much of the temperature range 16--500\,K.

It is an interesting and, as yet, unresolved issue as to how these large changes in framework geometry are accommodated at the atomic scale. The separations between some neighbouring pairs of atoms must be affected just as strongly by temperature as the overall lattice dimensions, and the important question is how these changes can be correlated throughout the crystal lattice with such a low energy penalty. This is essentially a \emph{local structure} problem, and as such one needs to employ experimental techniques capable of probing local correlations in order to address these issues. In previous studies of flexible materials, NTE phases and other disordered systems, we have seen that neutron total scattering can be an invaluable tool when studying local structure and dynamics \cite{Tucker_2005,Keen_2007,Tucker_2000,Tucker_2001,Goodwin_2007}. Indeed the technique is ideally suited to thermal expansion studies because it probes at once both average and local structure, and so can be used to produce atomic-level descriptions of the crystal structure that are entirely consistent with the known changes in average structure.

In this paper, we use a combination of neutron total scattering and reverse Monte Carlo (RMC) analysis \cite{Dove_2002,Tucker_2007} to study local structure in Ag$_3$[Co(CN)$_6$] over the temperature range 10--300\,K. Our interest is in understanding how the local arrangements of atoms in the material and their correlated behaviour vary as a function of temperature, with particular emphasis on: (i) thermal changes in the geometry of the Co--CN--Ag--NC--Co linkages, (ii) the extent to which transverse vibrational motion of these linkages helps moderate expansion along the $\langle101\rangle$ crystal axes, (iii) the rigidity (or otherwise) of [CoC$_6$] coordination octahedra, and hence whether so-called rigid unit modes (RUMs \cite{Hammonds_1996}) play an important role in the lattice dynamics, and (iv) evidence for the anharmonicity of Ag$\ldots$Ag interactions.

Our paper begins with a review of the crystal structure of Ag$_3$[Co(CN)$_6$] and its thermal expansion behaviour. We also include a brief discussion regarding bonding in the material, based largely on a parallel density functional theory (DFT) investigation by one of our groups \cite{Calleja_2008}. Section \ref{methods} then describes the various experimental and computational methods used in our study, including our methods of overcoming the severe anisotropic peak broadening effects observed in low-temperature diffraction patterns. Our experimental results and discussion are given in section \ref{results}, where we report the bond length and bond angle distributions extracted from our total scattering data, together with the results of both real- and reciprocal-space analyses of correlated behaviour in our RMC configurations. The implications of these experimental and computational results are discussed with particular reference to our previous study of colossal thermal expansion behaviour in Ag$_3$[Co(CN)$_6$] \cite{Goodwin_2008}.

\section{Structure and Bonding in Ag$_3$[Co(CN)$_6$]: A Review}\label{structurereview}

\subsection{Crystal structure}

\begin{figure}
\begin{indented}
\item[]\includegraphics{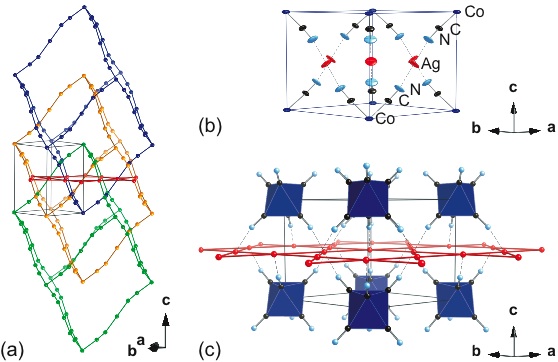}
\end{indented}
\caption{\label{fig1}Representations of the crystal structure of Ag$_3$[Co(CN)$_6$]. (a) Extended Co--CN--Ag--NC--Co linkages connect to form a three-dimensional framework whose topology is described by three interpenetrating cubic ($\alpha$-Po) nets. (b) The trigonal unit cell, with the cyanide linkages oriented along the $\langle101\rangle$ directions. (c) Elongation of the cubic nets along the trigonal axis produces a layered structure in which triangular arrays of hexacyanocobaltate anions (filled polyhedra) alternate with Ag-containing Kagome nets.}
\end{figure}

The first crystallographic study of Ag$_3$[Co(CN)$_6$] was reported in 1967 by Ludi and co-workers \cite{Ludi_1967}, but the validity of the structural model proposed in this paper was questioned in a subsequent re-evaluation by Pauling and Pauling \cite{Pauling_1968}. The Paulings' alternative solution, which was confirmed shortly afterwards \cite{Ludi_1968} and appears consistent with modern x-ray and neutron diffraction data \cite{Goodwin_2008}, describes a three-dimensional framework assembled from Co--CN--Ag--NC--Co linkages. Each Co$^{3+}$ centre is connected to six other Co atoms via these linkages, forming an extended covalent framework with a cubic network topology analogous to that of the Prussian Blues. The ``cube edges'' of the framework---namely the Co--CN--Ag--NC--Co linkages---are so long ($\simeq10$\,\AA) that two additional, identical framework lattices can be accommodated within the open cavities of the first, producing the triply-interpenetrating motif illustrated in Fig.~\ref{fig1}(a).

Threefold interpenetration of this type distinguishes one of the cube body diagonals from the remaining three, removing all but one of the original threefold axes of the cubic network. Consequently the highest compatible crystal symmetry is not a cubic space group, but rather the experimentally-observed trigonal group $P\bar31m$, with the $\bar3$ axis oriented parallel to the unique cube body diagonal. Consequently there is no symmetry constraint to prevent distortion of the cubic networks along the trigonal axis, and indeed at room temperature one observes a Co$\ldots$Co$\ldots$Co ``cube'' angle of 74.98$^\circ$ \cite{Goodwin_2008}. Inspection of the crystallographic unit cell determined from our recent neutron diffraction study [Fig.~\ref{fig1}(b)] suggests that this deviation from pseudo-cubic symmetry is accommodated by flexing of the Co--CN--Ag linkages (which now run parallel to the $\langle101\rangle$ axes of the trigonal lattice) rather than any significant distortion of the [CoC$_6$] coordination geometries: the C--Co--C angles deviate by less than 3$^\circ$ from the expected octahedral values \cite{Goodwin_2008}.  

By distorting in this way, the framework assumes a layered structure in which alternate sheets of Ag$^+$ and [Co(CN)$_6$]$^{3-}$ ions are stacked parallel to the unique axis of the trigonal cell [Fig.~\ref{fig1}(c)]. The silver atoms within a given layer lie at the vertices of a Kagome lattice, and the octahedral hexacyanocobaltate ions form a triangular lattice with vertices above and below the hexagonal Kagome ``holes''. This same layered motif is seen in the isostructural rare-earth salts La[Ag$_x$Au$_{1-x}$(CN)$_2$]$_3$ \cite{Colis_2005,Colis_2005b,Larochelle_2006}, Eu[Ag$_x$Au$_{1-x}$(CN)$_2$]$_3$ \cite{Colis_2005c} and L[M$^\prime$(CN)$_2$]$_3$ (L=Tb, Gd, Y; M$^\prime$ = Ag, Au) \cite{RawashdehOmary_2000}, and again in the chiral $P312$ transition-metal derivatives KCo[Au(CN)$_2$]$_3$ \cite{Abrahams_1980}, RbCd[Ag(CN)$_2$]$_3$ \cite{Hoskins_1994}, KFe[Au(CN)$_2$]$_3$ \cite{Dong_2003} and KMn[Ag(CN)$_2$]$_3$ \cite{Geiser_2003}. In this second group of compounds, the negative charge of the transition-metal-cyanide framework is counterbalanced by the inclusion of alkali earth cations within the framework cavities (the inclusion being responsible for the lower symmetry).

Amongst all these structurally-related materials, Ag$_3$[Co(CN)$_6$] is unique in terms of the orientation of the cyanide ions: the $d^{10}$ centres are coordinated by the N atoms of the cyanide groups. That we have a Co--CN--Ag arrangement, as opposed to Co--NC--Ag, is very unusual for an Ag-containing cyanide compound, which would normally be expected to contain the C-bound dicyanoargentate [Ag(CN)$_2$]$^-$ species \cite{Sharpe_1976,Geiser_2003}. The different arrangement seen here may be forced by the coordinatively-inert [Co(CN)$_6$]$^{3-}$ salt from which the material is prepared \cite{Goodwin_2008}. Perhaps a synthetic route involving Co$^{3+}$ and [Ag(CN)$_2$]$^-$ ions would yield the ``expected'' Co[Ag(CN)$_2$]$_3$ isomer; however, DFT calculations suggest that the Co--CN--Ag bonding arrangement is in fact thermodynamically favoured \cite{Calleja_2008}.

A second point of interest, and one that is discussed in more detail below, is the relatively short nearest-neighbour Ag$\ldots$Ag distances: the separation at room temperature is approximately 3.5\,\AA, which is only marginally greater than the van der Waals limit (3.4\,\AA\ \cite{Bondi_1964}), despite the coulombic repulsion. This close approach between $d^{10}$ centres is indicative of $d^{10}\ldots d^{10}$ ``metallophilic'' interactions \cite{Schmidbaur_2001,Pyykko_2004}. Our previous neutron scattering analysis \cite{Goodwin_2008} showed that anharmonicity of these interactions is implicated in the unusual thermal expansion behaviour of the material. Similarly, our separate DFT study \cite{Calleja_2008} showed the interactions also help ``soften'' the material by flattening the potential energy surface, facilitating the large changes in cell dimensions observed experimentally.

The crystal symmetry imposes a number of constraints on the framework geometry that restrict how it can respond to changes in temerapture. These can be summarised as follows:
\begin{enumerate}
\item{The Co atoms (which are all equivalent) have $\bar3m$ point symmetry, so that each [CoC$_6$] coordination octahedron may distort along the trigonal axis, but all Co--C bonds are equivalent and there are only three internal C--Co--C angles: $(90+\delta)^\circ$, $(90-\delta)^\circ$ and $180^\circ$.}
\item{The Ag atoms (which also are equivalent) are coplanar; with a single nearest-neighbour Ag distance and all nearest-neighbour Ag--Ag--Ag angles equal to $60^\circ$, $120^\circ$ or $180^\circ$.}
\item{The Ag site has $2/m$ point symmetry; consequently, all atoms within a given Co--CN--Ag--NC--Co linkage are coplanar, the N--Ag--N angle is equal to 180$^\circ$ and all cyanide groups are equivalent.}
\end{enumerate}

\subsection{Thermal expansion behaviour}

Our primary interest in Ag$_3$[Co(CN)$_6$] has been its highly unusual thermal expansion behaviour. When heated, the crystal lattice responds by expanding remarkably quickly along the $\mathbf a$ and $\mathbf b$ crystal axes, while contracting equally strongly along the $\mathbf c$ (trigonal) axis. In absolute terms, x-ray diffraction data show the unit cell parameters to vary between extremes of $a=6.7572(8), c=7.3731(13)$\,\AA\ at 16\,K and $a=7.1934(9), c=6.9284(15)$\,\AA\ at 496\,K \cite{Goodwin_2008}. Over much of this temperature range, the coefficients of thermal expansion are found to be essentially constant and equal to $\alpha_a=+132$\,MK$^{-1}$ and $\alpha_c=-130$\,MK$^{-1}$; these values correspond to a volume coefficient of thermal expansion of $\alpha_V=+134$\,MK$^{-1}$ and an equivalent isotropic expansion of $\alpha_\ell=+45$\,MK$^{-1}$.

For the vast majority of useable materials, uniaxial thermal expansivities lie in the range $0<\alpha<+20$\,MK$^{-1}$ \cite{Barron_1999,Krishnan_1979}, and so the behaviour exhibited by Ag$_3$[Co(CN)$_6$] is really rather exceptional. Indeed the positive thermal expansion (PTE) effect is similar in magnitude to that of the most weakly bound solids (\emph{e.g.}\ Xe \cite{Sears_1962}), while the NTE effect along $\mathbf c$ is unprecedented for a crystalline material.

The unusual behaviour of Ag$_3$[Co(CN)$_6$] appears to be intimately related to the flexibility of its framework lattice. In particular, the coupling between expansion along $\mathbf a$ and contraction along $\mathbf c$ means the overall behaviour can be interpreted as a variation of the Co$\ldots$Co$\ldots$Co ``cube'' angle [Fig.~\ref{fig1}(a)]. Because any such variation does not significantly affect the strongest bonding interactions---namely those along the Co--CN--Ag--NC--Co linkages---this process can occur with minimal cost in lattice enthalpy. What this means is that the absolute value of the Co$\ldots$Co$\ldots$Co angle at any given temperature (and hence the absolute unit cell dimensions) is essentially determined only by the weak Ag$\ldots$Ag interactions that occur \emph{between} interpenetrated networks (see below).

\subsection{Bonding}

As for all bridging transition-metal cyanides, one expects some degree of covalency in the Co$^{3+}\ldots$(CN)$^-$ and Ag$^+\ldots$(NC)$^-$ interactions. This expectation is consistent with the results of DFT calculations \cite{Calleja_2008}, which give Mulliken bond orders (overlap populations) of 0.25 (Co--C), 1.80 (C--N) and 0.34 (Ag--N) \cite{Segall_1996,Segall_1996b}. The difference in Co--C and Ag--N bond orders reflects the low electron density near the Co centre: the Co atom has a relatively high nominal charge, and at the same time the cyanide ligand is polarised such that the C atom is more electron-poor than the N atom. The same DFT calculations give Mulliken charges of $Q_{\rm Co}=+1.16$, $Q_{\rm Ag}=+0.65$, $Q_{\rm C}=-0.01$ and $Q_{\rm N}=-0.51$, which suggest there exists a substantial degree of charge transfer from the cyanide ligands to the cobalt atoms.

The covalency reflected in these DFT results is responsible for strengthening the bonding along the cube ``edges'' of the framework---the Co--CN--Ag--NC--Co linkages---and hence the coupling between thermal expansion behaviour along the $\mathbf a$ and $\mathbf c$ axes. It will also mean that deformation of the transition metal coordination geometries will carry a higher energy penalty than would otherwise be the case. In terms of the lattice dynamics, this would see phonon modes that preserve these coordination geometries (namely the RUMs) emerge as some of the lowest-energy lattice vibrations.

Our recent structural investigation suggested that the only significant bonding interactions \emph{between} interpenetrated frameworks are dispersion-like forces between neighbouring Ag atoms \cite{Goodwin_2008}. Rather than involving shared electrons, these ``argentophilic'' bonds arise from multipolar interactions between excited electronic states of the $d^{10}$ centres. As such, the interactions are not explicitly taken into account at the DFT level; indeed DFT gives the anticipated Ag$\ldots$Ag bond order of 0.0 (reflecting the absence of any covalency), together with a separation closer to that expected in the absence of any attractive terms \cite{Calleja_2008}. In fact the large difference between DFT and experimental cell parameters is actually strong evidence for the existence and dispersion-like nature of these interactions \cite{Calleja_2008}. It is difficult to estimate the strength of these Ag$\ldots$Ag interactions based on the DFT results alone; a comprehensive \emph{ab initio} study of [Cl--Ag--PH$_3$]$_2$ dimers suggests a bond energy of approximately 15\,kJmol$^{-1}$ \cite{OGrady_2004}, but larger values can also be found elsewhere in the literature \cite{Schmidbaur_2001,Pyykko_2004}. In either case, what is clear is that the energies involved are at least an order of magnitude lower than typical values for covalent or ionic bonds, and this results in a heightened sensitivity to the effects of \emph{e.g.}\ temperature and pressure.

Even without any explicit consideration of the dispersion-like Ag$\ldots$Ag interactions, DFT calculations show that the framework can adopt a wide variety of different geometries with essentially equivalent lattice enthalpies \cite{Calleja_2008}. For each geometry, the covalent Co--C, C--N and Ag--N interactions are unchanged (and the Ag$\ldots$Ag overlap populations remain equal to zero), but the Ag atoms show considerable tolerance to a range of Ag$\ldots$Ag distances. To some extent this will reflect the relatively high polarisability of the Ag$^+$ cation \cite{Iwadate_1982}, which translates to a ``fluffiness'' of its valence electron cloud. This ``fluffiness'' may well contribute to the superionic behaviour of some Ag$^+$ salts \cite{Hull_2004}, where cation polarisability can play a key role in diffusion mechanisms \cite{Castiglione_1999}. The relevance here is that this polarisability will facilitate flexibility within the Ag$_3$[Co(CN)$_6$] framework by allowing relatively large displacements of the Ag atoms as the material is heated.

\section{Methods}\label{methods}
\subsection{Neutron total scattering}

The neutron time-of-flight diffractometer GEM~\cite{Williams_1998,Day_2004,Hannon_2005} at the ISIS pulsed spallation source was used to collect total scattering patterns from a polycrystalline Ag$_3$[Co(CN)$_6$] sample, prepared as in \cite{Goodwin_2008}. Data were collected over a large range of scattering vectors of magnitudes $0.3\leq Q\leq50$\,\AA$^{-1}$, giving a real-space resolution of order $\Delta r\simeq3.791/Q_{\rm max}\simeq0.08$\,\AA. For the experiment, approximately 3\,g of sample was placed within a cylindrical thin-walled vanadium can of 8.3\,mm diameter and 5.8\,cm height, which was in turn mounted within a top-loading CCR. Data suitable for subsequent RMC analysis were collected at temperatures of 10, 50, 150 and 300\,K.

Following their collection, the total scattering data were corrected using standard methods, taking into account the effects of background scattering, absorption, multiple scattering within the sample, beam intensity variations, and the Placzek inelasticity correction \cite{Dove_2002}. These corrected data were then converted to experimental $F(Q)$ and $T(r)$ functions \cite{Dove_2002,Keen_2001}, which are related to the radial distribution function $G(r)$ by the equations
\begin{eqnarray}
F(Q)&=&\rho_0\int_0^\infty4\pi r^2G(r)\frac{\sin Qr}{Qr}{\rm d}r,\\
T(r)&=&4\pi r\rho_0\left[G(r)+\left(\textstyle\sum_mc_mb_m\right)^2\right],
\end{eqnarray}
where $\rho_0$ is the number density, $c_m$ the concentration of each species $m$ and $b_m$ the corresponding neutron scattering lengths. The Bragg profiles for each data set were extracted from the scattering 	data collected by the detector banks centred on scattering angles $2\theta=63.62$, $91.37$ and $154.46^\circ$.

\subsection{Average structure refinement}

\begin{figure}
\begin{indented}
\item[]\includegraphics{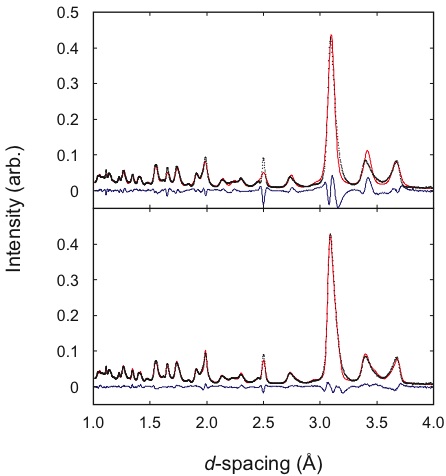}
\end{indented}
\caption{\label{fig2}Bragg profile refinements of Ag$_3$[Co(CN)$_6$] neutron diffraction data collected at 10\,K: (top) using a single-phase {\sc gsas} refinement with variable time-of-flight peak shape parameters, and (bottom) using our multiple-phase {\sc gsas} refinement and time-of-flight peak shape parameters obtained from data measured at room temperature. Experimental data are given as points, the fitted profile is shown as a solid line, and the difference (fit$-$data) is shown beneath each curve.}
\end{figure}

The experimental Bragg diffraction profiles were fitted with the {\sc gsas} Rietveld refinement program \cite{Larson_2000} using the published structural model \cite{Pauling_1968}. As we found in our previous crystallographic study \cite{Goodwin_2008}, the room temperature diffraction pattern could be fitted well using a standard peak shape function, while data collected at lower temperatures required special treatment [Fig.~\ref{fig2}]. 

In order to refine those data collected at lower temperatures, and their increased variation in peak widths, we considered that the polycrystalline sample contained a distribution of lattice parameters \cite{Stephens_1999}. Accordingly, we employed a multi-phase model in our structural refinements in which the Ag$_3$[Co(CN)$_6$] powder sample is treated as a mixture of phases with different unit cell parameters but with common relative atomic positions and atomic displacement parameters (at a given temperature). There is an implicit assumption here that the same set of atomic coordinates can be used for a distribution of cell parameters at a given temperature. While this will translate to some additional uncertainty in \emph{e.g.}\ bond lengths, we found the concession to be necessary in order to obtain fits to data that might be considered at all reasonable. For a given temperature, the contribution of each phase (calculated using the room temperature peak-shape parameters) is added to produce a composite diffraction profile that is capable of assuming anisotropic peak shapes. The {\sc gsas} refinement process then involved variation not only of the atomic positions and displacement parameters, but also of the relative populations of each phase. As such, we avoided refining a single set of lattice parameters, obtaining instead a distribution profile across a range of unit cell dimensions.

For each data set, we refined the populations of eight separate phases, choosing lattice parameters for each phase according to the following algorithm, developed in order to model the observed anisotropy most satisfactorily. First, a structure refinement was performed using a standard t.o.f.\ peak shape function, giving a ``compromise'' set of lattice parameters $a_{\rm n},c_{\rm n}$ that resemble a weighted mean across the entire distribution. In each case, these values reflected a more moderate thermal expansion behaviour than those values $a_{\rm x},c_{\rm x}$ obtained in the {\sc topas} \cite{Coelho_2000} refinements of x-ray data published elsewhere \cite{Goodwin_2008} (\emph{i.e.}\ $a_{\rm x}<a_{\rm n}$ and $c_{\rm x}>c_{\rm n}$). We note that this is to be expected since the x-ray values correspond to the single-crystal limit; \emph{i.e.}, the values expected in the absence of strain effects. The differences $\delta_a=a_{\rm x}-a_{\rm n}$ and $\delta_c=c_{\rm x}-c_{\rm n}$ between these two lattice parameters are then used to produce a set of eight lattice parameters for the {\sc gsas} multi-phase refinement: $\{(a_{\rm n}+\epsilon\delta_a,c_{\rm n}+\epsilon\delta_c); \epsilon=\frac{4}{3},\frac{2}{3},0,-\frac{2}{3},-\frac{4}{3},-\frac{6}{3},-\frac{8}{3},-\frac{10}{3}\}$. This range of $\epsilon$ values was judged the minimum necessary to model the peak asymmetry. The values of $a_{\rm x}$, $a_{\rm n}$, $c_{\rm x}$ and $c_{\rm n}$ and the resultant $\delta_a$, $\delta_c$ are listed in Table~\ref{table_gsasparams}; the refined populations for different values of $\epsilon$ are given in Table~\ref{table_epsilons}. Because of the relatively small values for $\delta_a$ and $\delta_c$ at 300\,K, we used only a single-phase refinement (with lattice parameters $a_{\rm n}$ and $c_{\rm n}$) for this data set.

\begin{table}
\caption{\label{table_gsasparams} Lattice parameter values $a,c$ refined from neutron (subscript ``n'') and x-ray (subscript ``x'') \cite{Goodwin_2008} diffraction patterns, together with their differences $\delta$. The neutron-derived values are ``compromise'' values, obtained using a standard time-of-flight peak shape that does not account for the distribution in lattice parameters observed at low temperatures [\emph{cf.}\ the top panel in Fig.~\ref{fig2} and see text for further details].}
\begin{center}
\begin{tabular}{ccccccc}      
\hline\hline $T$/K &$a_{\rm n}$/\AA&$a_{\rm x}$/\AA&$\delta_a$/\AA&$c_{\rm n}$/\AA&$c_{\rm x}$/\AA&$\delta_c$/\AA\\\hline
10&6.8256(5)&6.7537(8)&0.0719(13)&7.3294(11)&7.3806(13)&$-$0.0512(24)\\
50&6.8485(5)&6.7805(7)&0.0680(12)&7.3052(10)&7.3522(12)&$-$0.0470(22)\\
150&6.9239(3)&6.8879(5)&0.0360(8)&7.2307(6)&7.2587(9)&$-$0.0281(15)\\
300&7.03066(9)&7.0255(5)&0.0052(6)&7.11748(17)&7.1251(11)&$-$0.0076(13)\\
\hline
\end{tabular}
\end{center}
\end{table}

\begin{table}
\caption{\label{table_epsilons} Lattice parameter distribution populations refined using our multiple-phase {\sc gsas} treatment of the low-temperature anisotropic peak broadening observed in neutron diffraction patterns. A single coefficient vanished systematically during refinement and so its value was constrained to be equal to zero (shown in italics). See text for further details.}
\begin{center}
\begin{tabular}{cccc}      
\hline\hline $\epsilon$&\multicolumn{3}{c}{$T$/K}\\
\cline{2-4}\\[-10pt]
&10&50&150\\[2pt]\hline
$\frac{4}{3}$&0.1281(21)&0.1360(17)&0.2610(21)\\
$\frac{2}{3}$&0.2785(23)&0.2386(10)&0.1628(16)\\
$0$&0.2147(24)&0.2209(17)&0.2105(23)\\
$-\frac{2}{3}$&0.1207(24)&0.1555(12)&0.0836(18)\\
$-\frac{4}{3}$&0.0956(26)&0.0931(20)&0.106(3)\\
$-\frac{6}{3}$&0.0845(26)&0.0739(20)&0.047(3)\\
$-\frac{8}{3}$&0.0332(26)&0.0312(20)&\emph{0}\\
$-\frac{10}{3}$&0.0447(21)&0.0509(17)&0.1288(20)\\
\hline
\end{tabular}
\end{center}
\end{table}

\subsection{Reverse Monte Carlo refinement}

The reverse Monte Carlo refinement method as applied to crystalline materials, together with its implementation in the program {\sc rmcp}rofile have been described in detail elsewhere \cite{Dove_2002,Tucker_2007}. The basic refinement objective is to produce atomistic configurations that account simultaneously for the experimental $F(Q)$, $T(r)$ and Bragg profile $I(t)$ functions. This is achieved by accepting or rejecting random atomic moves subject to the metropolis Monte Carlo algorithm, where in this case the Monte Carlo ``energy'' function is determined by the quality of the fits to data.

In light of the anisotropic peak broadening effects at low temperatures, we used a modified version of the {\sc rmcp}rofile code that was capable of taking into account the lattice parameter distributions described above. The experimental data were compared against an appropriately weighted sum of individual $F(Q)$, $T(r)$ and $I(t)$ functions, each corresponding to a different set of lattice parameters but calculated from a common atomistic configuration. This approach was deemed necessary because anisotropic variation of the lattice parameters affects the powder-averaged pair distribution function in a non-trivial manner.

Our starting configurations for the RMC process were based not on the crystallographic unit cell shown in Fig.~\ref{fig1}(b), but on a supercell with orthogonal axes given by the transformation
\begin{equation}
\left[\begin{array}{l}\mathbf a\\ \mathbf b\\ \mathbf c\end{array}\right]_{\rm RMC}=\left[\begin{array}{ccc}6&0&0\\ -4&8&0\\ 0&0&6\end{array}\right]\times\left[\begin{array}{l}\mathbf a\\ \mathbf b\\ \mathbf c\end{array}\right]_{P\bar31m}.
\end{equation}
The use of orthogonal axes facilitates the preparation of RMC configurations that are approximately the same length in each direction, maximising the pair distribution cut-off value $r_{\rm max}$ for a given number of atoms. Our configurations contained 4608 atoms and extended approximately 42\,\AA\ in each direction. Prior to RMC refinement, small random initial displacements were applied to each atom in the configuration, and a set of data-based ``distance window'' constraints were set in place in order to maintain an appropriate framework connectivity throughout the refinement process \cite{Tucker_2007,Goodwin_2005c}. The values used for these constraints are given in Table~\ref{table1}, where they are shown not to interfere with the relevant bond-length distributions. We chose not to include ``soft'' bond-length or bond-angle restraints in the refinement, in order to avoid the incorporation of any dynamical bias \cite{Goodwin_2005c}.

\begin{table}
\caption{\label{table1} ``Distance window'' parameters $d_{\rm min},d_{\rm max}$ used in our Ag$_3$[Co(CN)$_6$] RMC refinements  and the corresponding mean pair separations $\bar d$ and their standard deviations $\sigma$ at 300\,K (where the distributions are broadest).}
\begin{center}
\begin{tabular}{ccccc}      
\hline\hline Atom pair&$d_{\rm min}$/\AA&$d_{\rm max}$/\AA&$\bar d$/\AA&$\sigma$/\AA\\\hline
Co--C&1.76&2.22&1.928&0.085\\
C--C&2.44&3.00&2.718&0.135\\
C--N&1.00&1.29&1.161&0.055\\
Ag--N&1.76&2.22&2.084&0.070\\
Ag--Ag&3.00&4.10&3.519&0.156\\\hline
\end{tabular}
\end{center}
\end{table}

\begin{figure}
\begin{indented}
\item[]\includegraphics{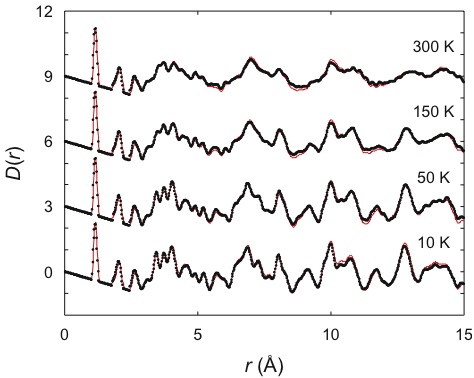}
\end{indented}
\caption{\label{fig3}Experimental $D(r)$ data (points) and RMC fits (solid lines) obtained using {\sc rmcp}rofile as described in the text. Data for successive temperature points have been shifted vertically by three units in each case.}
\end{figure}

\begin{figure}
\begin{indented}
\item[]\includegraphics[width=10cm]{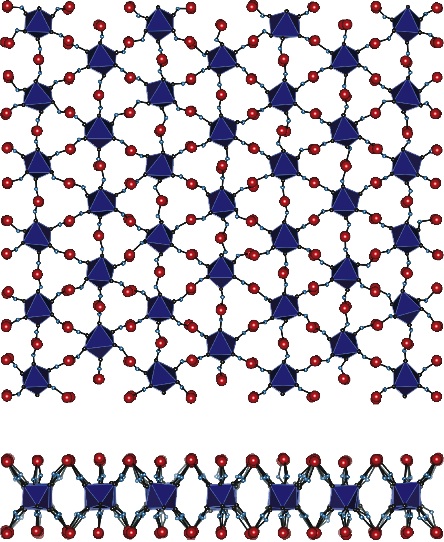}
\end{indented}
\caption{\label{fig4}A $(001)$ slice of a 300\,K RMC configuration, viewed along a direction parallel (top) and perpendicular (bottom) to the trigonal axis $\mathbf c$.}
\end{figure}

The refinement process was allowed to continue until no further improvements in the fits to data were observed. The real-space fits obtained for each of the four temperature points are illustrated in Fig.~\ref{fig3}, where we have converted the experimental and modelled $T(r)$ functions to the related functions
\begin{equation}
D(r)=T(r)-4\pi r\rho_0\left(\textstyle\sum_mc_mb_m\right)^2=4\pi r\rho_0G(r)
\end{equation}
for ease of representation \cite{Dove_2002,Keen_2001}. The $F(Q)$ and Bragg profile $I(t)$ functions were modelled similarly successfully, and the quality of these fits was just as consistent across the four data sets. A final check of the integrity of the RMC models is to determine whether any anomalous regions of ``damage'' or unphysical atomic displacements have evolved during the refinement process, as discussed in more detail elsewhere with respect to RMC refinements of SrTiO$_3$ \cite{Goodwin_2005c}. Here, visual inspection of our RMC configurations confirms the absence of any such regions; a representative section of a 300\,K RMC configuration is illustrated in Fig.~\ref{fig4}. Furthermore, to ensure reproducibility in our results, all positional, bond length and bond angle distributions presented in our analysis were calculated as averages over 10 independent RMC refinements.

\subsection{Phonon analysis}

The {\sc rmcp}rofile implementation of the reverse Monte Carlo (RMC) structural refinement method \cite{Tucker_2007}, modified as described above, was used to generate ensembles of atomistic configurations consistent with the neutron $T(r)$, $F(Q)$ and $I(t)$ data. Each ensemble (corresponding to a single temperature point) contained approximately 1000 independent configurations. The atomic positions in these ensembles were analysed via the reciprocal-space approach given in \cite{Goodwin_2005c,Goodwin_2004} to yield a set of phonon frequencies and mode eigenvectors for each point in reciprocal space allowed by the supercell geometry. Essentially what this process does is to translate the magnitude of correlated atomic displacements into the energy of lattice vibrations: larger displacements will correspond to lower phonon energies, and smaller displacements to higher energies.

There is a difficulty in comparing the phonon frequencies obtained for different temperature points in that RMC configurations will contain a level of baseline ``noise'' that arises because, amongst other effects, instrument resolution is not explicitly taken into account. This has the effect of making the displacements at low temperatures appear to be much greater than one would expect, while those at higher temperatures (where the ``noise'' becomes less significant with respect to the ``true'' correlated motion) become increasingly accurate. Consequently, one observes a systematic shift in phonon frequencies for variable-temperature studies that can make quantitative analysis difficult \cite{Goodwin_2005c}. Here we have scaled the phonon frequencies at temperatures below 300\,K such that they give the same mean phonon frequency as observed at 300\,K itself (where we expect this value to be most robust). This means that changes in the total phonon energy are not recoverable from the analysis, but we can still see how the energies of partial densities of states, \emph{i.e.}\ individual components of the phonon spectrum, are affected by temperature \emph{relative to one another}. 

\section{Results and Discussion}\label{results}
\subsection{Average structure}

By analysing our neutron scattering data using both {\sc gsas} and {\sc rmcp}rofile approaches, we obtain two subtly different, but complementary, descriptions of the average structure (\emph{i.e.}\ the distribution of atomic positions). On the one hand, {\sc gsas} refinements give accurate measurements of the lattice parameters and the average positions of atoms in the unit cell. On the other hand, {\sc rmcp}rofile gives a three-dimensional positional distribution function for each atom: it models thermal displacements as a distribution of partial atom occupancies rather than as ellipsoidal Gaussian functions.

\begin{figure}
\begin{indented}
\item[]\includegraphics{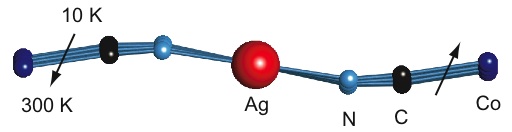}
\end{indented}
\caption{\label{fig5}Variation in the average geometry of the Co--CN--Ag--NC--Co linkage in Ag$_3$[Co(CN)$_6$] over the temperature range 10--300\,K as determined by {\sc gsas} refinements of Bragg profile data.}
\end{figure}

The average atomic coordinates refined using {\sc gsas} are given in Table~\ref{avgcoords}. Perhaps the most interesting result from these refinements is the variation in average geometry of the Co--CN--Ag--NC--Co linkage as a function of temperature, illustrated in Fig.~\ref{fig5}. What is apparent from this figure is that the lattice parameter variations occur with minimal effect on the Co--C and C--N bonding geometries. Instead, the C--N--Ag angle increases with increasing temperature, such that the linkage approaches a more linear average coordination geometry. We note that the same qualitative behaviour was observed in our parallel DFT study \cite{Calleja_2008}. Because of these geometric changes in the Co--CN--Ag--NC--Co linkage, the actual thermal expansion behaviour of the Co$\ldots$Co vector is relatively complicated. The move towards a more linear geometry at higher temperatures will produce a PTE effect by pushing the Co atoms apart as the linkage straightens. However, we would also expect transverse vibrational motion of the C, N and Ag atoms to draw the Co ends closer together at the same time \cite{Goodwin_2006}. What happens in practice---illustrated most clearly in our previous x-ray diffraction results \cite{Goodwin_2008}---is that the expansion behaviour is very much weaker than the colossal effects along the crystal axis, and there is also a subtle alternation between low-level PTE and NTE behaviour, which is entirely compatible with the existence of two competing effects.

\begin{table}
\caption{\label{avgcoords}Relative average atomic coordinates from multiple-phase {\sc gsas} and {\sc rmcp}rofile refinements of Ag$_3$[Co(CN)$_6$]: Co at $(0,0,0)$, Ag at $(\frac{1}{2},0,\frac{1}{2})$, C at $(x_{\rm C},0,z_{\rm C})$ and N at $(x_{\rm N},0,z_{\rm N})$.}
\begin{center}
\begin{tabular}{lccccc}      
\hline\hline &$T$/K&$x_{\rm C}$&$z_{\rm C}$&$x_{\rm N}$&$z_{\rm N}$\\\hline
{\sc gsas}&10&0.21994(26)&0.15280(23)&0.34167(20)&0.26553(15)\\
&50&0.21968(26)&0.15301(23)&0.34033(20)&0.26589(15)\\
&150&0.21376(25)&0.15742(27)&0.33778(20)&0.26653(17)\\
&300&0.21160(21)&0.16080(19)&0.33377(13)&0.26737(17)\\\hline
{\sc rmcp}rofile&10&0.22195(12)&0.15587(13)&0.33855(17)&0.26443(12)\\
&50&0.22256(13)&0.15559(14)&0.33784(18)&0.26475(13)\\
&150&0.22083(8)&0.15683(13)&0.33498(15)&0.26517(13)\\
&300&0.21993(16)&0.15712(15)&0.33089(24)&0.26644(18)\\\hline
\end{tabular}
\end{center}
\end{table}

\begin{figure}
\begin{indented}
\item[]\includegraphics{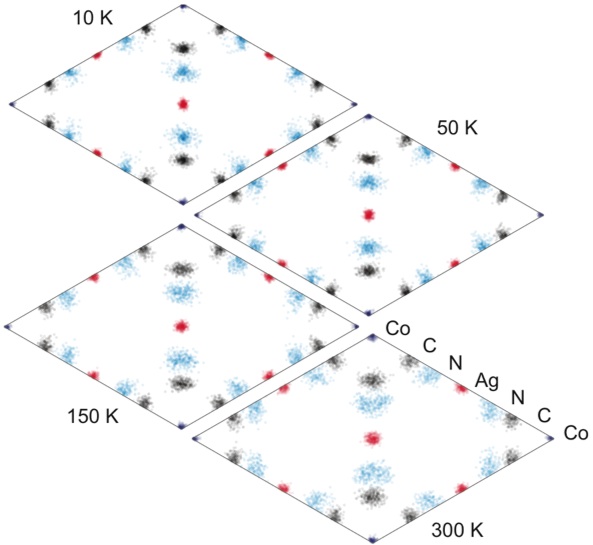}
\end{indented}
\caption{\label{fig6}Atomic distributions obtained by collapsing RMC configurations onto a single unit cell, viewed down the crystallographic $\mathbf c$ axis.}
\end{figure}

The atomic distributions calculated by collapsing representative RMC configurations onto a single unit cell are shown in Fig.~\ref{fig6}. These illustrate the absence of any split-site disorder, with thermal displacements centred on the average crystallographic positions. As expected, these displacements are rather anisotropic in the case of C and N atoms (and, to a lesser extent, for the Ag atoms too). The distributions at low temperatures are perhaps a little broader than expected; the origin of this broadening is understood \cite{broadening_note} and is seen in other RMC studies (\emph{e.g.}\ \cite{Goodwin_2005c}). The average atomic coordinates obtained using {\sc rmcp}rofile are given in Table~\ref{avgcoords}, where they are compared with the values obtained from {\sc gsas} refinements. The RMC values show a smoother variation with respect to temperature than the {\sc gsas} results. However the general changes in structure, such as flexing of the Co--CN--Ag--NC--Co linkages as illustrated in Fig.~\ref{fig5}, remain consistent between the two approaches.

\subsection{Local structure: bond length and bond angle distributions}

The ``bond lengths'' calculated from average structure determinations such as Rietveld refinement really only represent the separations between average atomic positions, denoted by the nomenclature $\langle$A$\rangle$--$\langle$B$\rangle$. RMC refinement of total scattering data allows us to measure the true average bond lengths $\langle$A--B$\rangle$ because these correspond directly to peak positions in the experimental $T(r)$ function.

\begin{figure}
\begin{indented}
\item[]\includegraphics{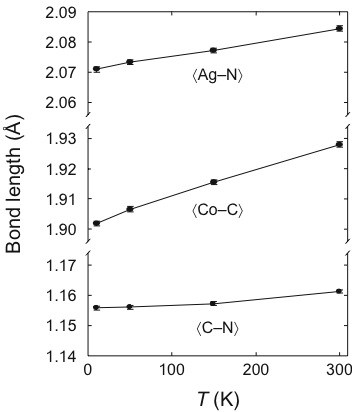}
\end{indented}
\caption{\label{fig7}Thermal variation in the average $\langle$C--N$\rangle$, $\langle$Co--C$\rangle$ and $\langle$Ag--N$\rangle$ bond lengths as given by the corresponding RMC pair distribution functions.}
\end{figure}

In this respect, the Co--C, C--N and Ag--N bond lengths are of particular interest because transverse vibrational motion of the C and N atoms means that the true distances cannot easily be determined from average structure analysis. This is especially difficult here because the spread in lattice parameters at low temperatures gives rise to a corresponding additional uncertainty in the bond lengths; indeed the scatter observed in our {\sc gsas} analysis is too large to consider the absolute values meaningful. On the other hand, the true average bond lengths $\langle$Co--C$\rangle$, $\langle$C--N$\rangle$ and $\langle$Ag--N$\rangle$ determined from our RMC models behave smoothly with respect to temperature [Fig.~\ref{fig7}]. What is clear is that all bond lengths increase linearly across the temperature range studied. The rate of increase of the Co--C separation approximately double that of the Ag--N separation, which in turn increases more strongly than the C--N separation.

The magnitude of the PTE effect observed for the Co--C and Ag--N bonds is perhaps surprising: the corresponding coefficients of thermal expansion are approximately equal to +45 and +25\,MK$^{-1}$, respectively. By way of a comparison, the coefficient of thermal expansion for the Zn--C/N bond in Zn(CN)$_2$ was found to be equal to +10.2(10)\,MK$^{-1}$ in an x-ray total scattering study \cite{Chapman_2005}. The accuracy of our values will be affected by the overlap between the Co--C and Ag--N distribution functions; we note also that the magnitude of the changes is small with respect to the real-space resolution given by $2\pi/Q_{\rm max}$.  A direct fit of two appropriately-weighted Gaussian distributions to the observed $T(r)$ data gives a similar value for expansion of the Ag--N bond (which is more strongly weighted in the neutron data), but a reduced expansivity for the Co--C bond of \emph{ca} +10\,MK$^{-1}$. 

\begin{figure}
\begin{indented}
\item[]
\includegraphics{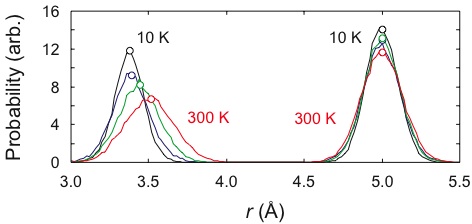}
\end{indented}
\caption{\label{fig8}Ag$\ldots$Ag (left) and Co$\ldots$Ag (right) nearest-neighbour pair distribution functions extracted from our RMC configurations. In each case, the broader distributions correspond to higher temperatures. The values determined from {\sc gsas} refinement of the average structure are given as open circles.}
\end{figure}

The changes in the Ag$\ldots$Ag distribution function are fundamentally different to the behaviour along the Co--CN--Ag linkage: the essentially-Gaussian distribution moves to higher $r$ values with increasing temperature as quickly as its breadth increases [Fig.~\ref{fig8}]. By comparison, the variations in bond lengths seen in Fig.~\ref{fig7} are all small with respect to the breadths of the corresponding partial pair distribution functions. It is important to keep in mind that the distributions at low temperature are artificially broadened as discussed above; consequently the ``true'' 10\,K Ag$\ldots$Ag distribution function would be much sharper than that shown in Fig.~\ref{fig8}. However the precision with which we can determine the mid-point of a distribution function is much smaller than the width of the distribution, and so the mean values are still well-constrained by the $T(r)$ data. In the harmonic limit, one expects these distribution functions to remain centred on the same value of $r$, and to broaden symmetrically with increasing temperature; here the behaviour reflects the anharmonicity of nearest-neighbour Ag$\ldots$Ag bonding interactions.

Conversely, the Co$\ldots$Ag distribution function shows that the separation between Co and Ag atoms remains essentially constant with respect to temperature, with the only noticeable thermal effect being an increased broadening at higher temperatures. Indeed the variation in mean values is found to be less than $0.005$\,\AA, which is similar to the changes in C--N bond lengths but substantially smaller than the increase in Co--C and Ag--N distances along the Co--CN--Ag linkages. This discrepancy can only be understood by allowing for a strong NTE effect of transverse vibrational motion of the C and N atoms.

\begin{figure}
\begin{indented}
\item[]
\includegraphics{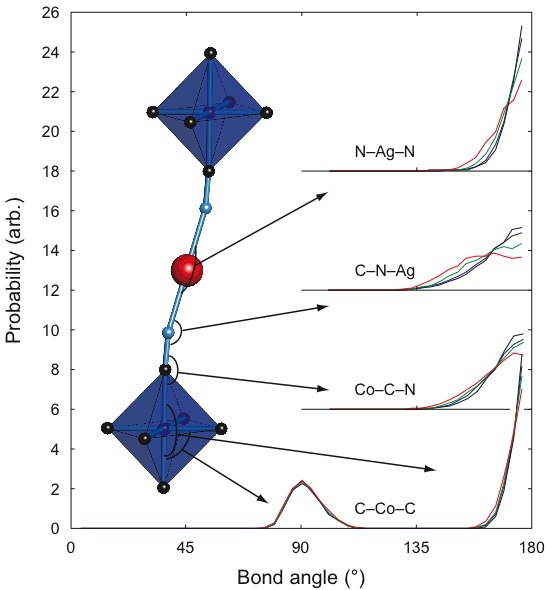}
\end{indented}
\caption{\label{fig9}RMC bond angle distributions: intra-octahedral C--Co--C angles and Co--C--N, C--N--Ag and N--Ag--N angles. Successive data sets are shifted vertically by multiples of 6 units. In all cases, the broader distributions correspond to higher temperatures.}
\end{figure}

Various bond angle distributions are shown in Fig.~\ref{fig9}, where they are given in a form that takes into account the fact that the number of angles around any particular value $\theta$ is proportional to $\sin\theta$. Considering first the intra-octahedral C--Co--C angles, we find that these behave much as might be anticipated in that the distributions, centred around 90$^\circ$ and 180$^\circ$, broaden very slowly with increasing temperature. There are two crystallographically-distinct C--Co--C angles near 90$^\circ$ (one corresponding to pairs of C atoms on the same side of the $z=0$ plane, and one to pairs of C atoms on opposite sides), and a slight difference in the breadths of the corresponding distributions may be responsible for the asymmetric peak shape observed. The fact that the experimental distribution functions do not change significantly across our different RMC refinements strongly indicates a relatively constrained [CoC$_6$] coordination geometry that is not easily affected by temperature.

We find much greater variation in bond angles along the Co--CN--Ag--NC--Co linkages. There are three symmetry-independent angles to consider---namely, Co--C--N, C--N--Ag and N--Ag--N---and the corresponding distributions are also shown in Fig.~\ref{fig9}. Again, the lowest-temperature distributions will be artificially broadened, and keeping this in mind we see that all three angles reflect a significant flexibility along the entire linkage. The strongest changes are observed for the C--N--Ag angles. This is entirely consistent with the conclusions drawn from average structure analysis, where we expected the large changes in framework geometry to be accommodated primarily via hingeing of the linear linkages at the N atom [Fig.~\ref{fig5}]. As anticipated for a linear coordination geometry, the distribution in N--Ag--N angles shows a much stronger thermal variation than the intra-polyhedral [CoC$_6$] angles.

That the C--Co--C angle distributions are essentially temperature-independent despite the large C atom displacements seen in the average structure analysis is a clear indication that the atomic displacements in our RMC configurations are strongly correlated. However, it is difficult to obtain a quantitative handle on these correlations from a consideration of pair and triplet distribution functions alone. Instead we turn to geometric algebra, which permits a more sophisticated real-space analysis.

\subsection{Geometric algebra analysis: Rigid unit modes}

By comparing the orientations of [CoC$_6$] octahedra within an ensemble of RMC configurations, it is possible to quantify two aspects of their dynamical behaviour: first, we obtain a measure of the magnitude of octahedral tilts for each octahedron; second, we can determine the extent to which each octahedron distorts via stretching of the Co--C bonds and/or bending of the C--Co--C angles. We carry out these calculations using a method based on geometric algebra \cite{Wells_2002}, as implemented in the program {\sc gasp} \cite{Wells_2004}. For each octahedron, the program first computes an associated ``rotor'', an algebraic quantity whose components represent, to first order, the degree of rotation about the three cartesian axes. These rotor components are then used by the program to determine the magnitudes of octahedral translations, rotations and distortions (due both to bending and to stretching of bonds). By analysing how these values are affected by temperature, we can determine whether the dominant lattice vibrations involve octahedral deformation modes, or alternatively whether the Co coordination geometries are reasonably well preserved. The individual rotor components also indicate whether any particular rotation directions are preferred.

\begin{figure}
\begin{indented}
\item[]
\includegraphics{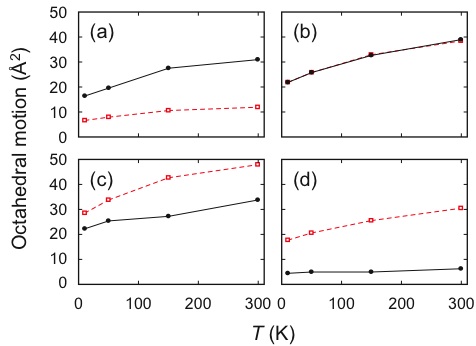}
\end{indented}
\caption{\label{fig10}[CoC$_6$] octahedral displacement components calculated using {\sc gasp} for our RMC configurations (filled circles, solid lines) and for ``reference'' configurations (open squares, dashed lines): (a) rotations, (b) translations, (c) bond-bending deformations, and (d) bond-stretching deformations.}
\end{figure}

The rotational, translational and distortive (C--Co--C bending and Co--C stretching) components of the octahedral motion calculated from our RMC configurations are illustrated in Fig.~\ref{fig10}. In this figure, the various components are compared with the values calculated from reference configurations, which share the same set of atomic displacements given by RMC but where these have now been distributed randomly throughout the configuration. Both RMC and ``reference'' configurations share the same average structure, but higher-order correlations will be constrained in the former via refinement against the experimental $T(r)$ function. This comparison then helps illustrate the way in which $T(r)$ data provide information about local correlations. In this specific case, it is apparent that the extent to which [CoC$_6$] octahedra are distorted by bond stretching and bond bending displacements is substantially smaller in our RMC configurations than is required by the average structure information. Instead, the atomic displacements appear to be correlated more strongly in the form of octahedral rotations, which occur with a significantly greater probability than the average structure demands. This is strong evidence of the RUM-type vibrations discussed above. It is straightforward to show that one expects the same degree of polyhedral translation in both cases (since this depends essentially on the root-mean-squared Co displacement, which is preserved), and that this is indeed the case is a good internal check on our analysis.

\begin{figure}
\begin{indented}
\item[]
\includegraphics{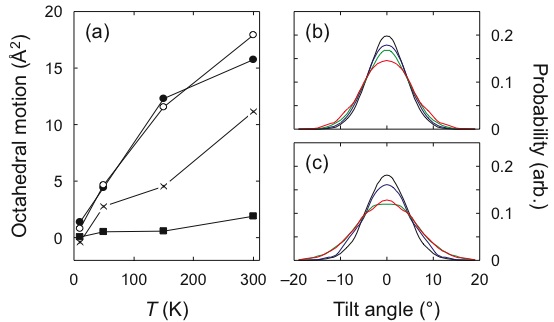}
\end{indented}
\caption{\label{fig11}(a) Thermal variation in [CoC$_6$] octahedral displacement patterns, after correction for the 0\,K RMC background: translations (open circles), rotations (filled circles), bond-bending deformations (crosses) and bond-stretching deformations (filled squares). The right-hand panels show the distributions of [CoC$_6$] rotation angles around (b) axes perpendicular to $\mathbf c$ and (c) an axis parallel to $\mathbf c$. In panels (b) and (c), the broader distributions correspond to higher temperatures.}
\end{figure}

As explained above our RMC configurations are known to contain a certain degree of random ``noise'', responsible for adding a baseline level of atomic motion that is particularly noticeable at low temperatures. Together with the baseline contribution due to zero point motion, these effects explain why our geometric algebra analysis shows non-zero levels of polyhedral motion and deformation even at low temperatures. By subtracting the projected 0\,K values, we arrive at the baseline-corrected plot shown in Fig.~\ref{fig11}(a). This representation of our {\sc gasp} analysis shows that on heating Ag$_3$[Co(CN)$_6$], one populates vibrational modes that involve translations and rotations of [CoC$_6$] in preference to C--Co--C bending modes and Co--C stretching modes. This is all very intuitive, and certainly consistent with the expectations of a RUM analysis \cite{Goodwin_2006}.

As discussed above, it is possible to determine the extent to which [CoC$_6$] octahedra rotate in different directions. By virtue of the $\bar3m$ site symmetry at the Co position, there are two key axes to consider: these correspond to rotations around $\mathbf a$ and others around $\mathbf c$. Panels (b) and (c) of Fig.~\ref{fig11} give plots of the distribution of rotation angles around both of these axes. This figure shows that the distribution around $\mathbf c$ is broader at each temperature point than that around $\mathbf a$, and hence the root-mean-squared deviation is larger, but that these axial rotations may have saturated by 150\,K. We note that with only one temperature point above 150\,K it is impossible for us to determine whether this saturation is indeed a real effect.

\subsection{Phonon analysis}

With 16 atoms in the primitive unit cell, the lattice dynamics of Ag$_3$[Co(CN)$_6$] is characterised by a set of 48 phonon modes. As such, this is a substantially more complex system than either MgO or SrTiO$_3$ (6 and 15 modes, respectively), for which a phonons-from-total-scattering (PFTS) approach has been applied previously \cite{Goodwin_2005c,Goodwin_2004}. The overall range in phonon frequencies is similar in all cases, and so an increase in the number of phonon modes corresponds to a finer separation between mode frequencies across the phonon spectrum. In terms of our phonon calculations this makes meaningful analysis of the three-dimensional RMC phonon dispersion curves particularly difficult, precisely because the confidence in mode frequency values approaches the separation between successive modes. Consequently, our PFTS analysis focusses on the phonon density of states, which helps condense the quantity of vibrational information into a more manageable form. 

\begin{figure}
\begin{indented}
\item[]
\includegraphics{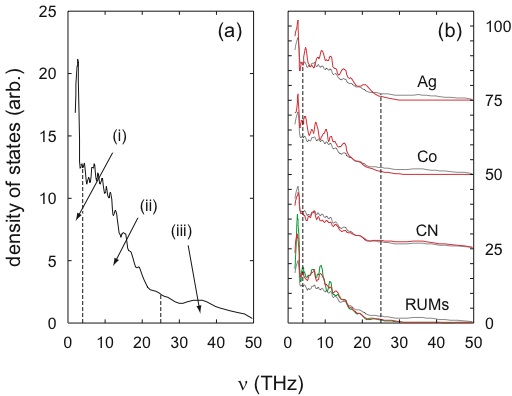}
\end{indented}
\caption{\label{fig12}(a) Powder-averaged phonon density of states calculated from an ensemble of 1000 separate 300\,K RMC configurations. The three energy regimes referred to in the text are labelled. (b) Scaled partial phonon densities of states, shifted vertically by multiples of 25 units. Each scaled partial density of states (thick line) is superposed on the overall density of states (thin line) for comparison. For the RUM curve, the component due to rotations around $\mathbf c$, which oscillates more strongly than the overall RMC curve, is also shown.}
\end{figure}

Considering first the overall structure of the phonon density of states, we obtain the most accurate picture from the ensemble of RMC configurations that corresponds to a temperature of 300\,K, where the effects of anisotropic peak broadening are negligible. The overall density of states reveals three main features: (i) a sharp peak at approximately 3\,THz; (ii) a broader distribution of modes, peaking at approximately 10\,THz and with an upper limit of \emph{ca} 25\,THz; and (iii) a very high frequency component, composed of a peak at 35\,THz together with a broader distribution centred around the same value [Fig.~\ref{fig12}(a)]. Random ``noise'' in the configuration has the effect of adding an exponential decay function to the PFTS density of states, and this is why the form of the d.o.s. function appears perhaps slightly unfamiliar. What we are really interested in is the deviation from this exponential background, which reflects the energies of correlated atomic displacements.

In order to determine the types of atomic displacements characterised by the different frequency regimes of this density of states, we calculated a series of partial phonon densities of states for different types of displacement. We calculated these partial densities of states by weighting each mode by the projection of its eigenvector onto a relevant set of displacement patterns. So, for example, a mode whose eigenvector could be represented as a linear combination of the RUMs would project strongly onto the set of RUM eigenvectors, and so its contribution to the RUM partial density of states would be large. The corresponding functions are shown in Fig.~\ref{fig12}(b) for Co atoms, Ag atoms, CN groups and also for RUM-type rotations of [CoC$_6$] octahedra (essentially a subset of the CN group displacements). Naturally each of these types of displacements contribute to modes at many different wave-vectors, but some overall features do emerge. What we are comparing at each stage is the likelihood of observing particular displacement patterns at a given energy value relative to the overall density of phonon modes at that energy.

Our analysis shows that the peak at lowest frequencies projects most strongly onto the RUM displacements, and especially those corresponding to rotations around $\mathbf c$. This reflects some of the results of the {\sc gasp} analysis above: the population of RUMs at low temperatures, and also the broader distributions for rotations around $\mathbf c$ than around $\mathbf a$.

The mid-frequency region contains components of all types of displacements, but the contribution of RUMs falls to zero most quickly on moving to higher energies.

The highest-frequency regime corresponds exclusively to stretching modes of the CN groups. The energy calculated for these modes comes directly from the width of the CN peak at 1.15\,\AA\ in the experimental $T(r)$ \cite{CN_peak_note}. This peak width is especially sensitive to the effects of truncation ripples in the Fourier transform from $F(Q)$, and so we expect the corresponding phonon frequencies to be substantially lower than experimental values. Here we obtain a mean value of approximately 35\,THz, which underestimates the literature values by around 10\,THz \cite{Sharpe_1976}. This difference corresponds to a broadening of the CN peak of less than 30\%, which is not at all unreasonable.

\begin{figure}
\begin{indented}
\item[]
\includegraphics{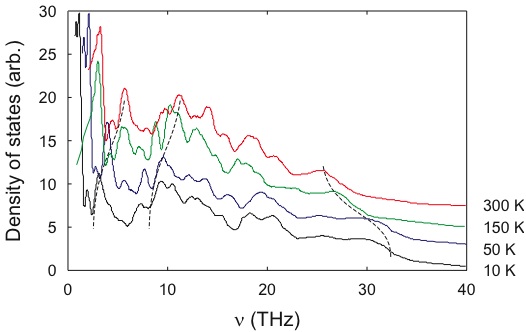}
\end{indented}
\caption{\label{fig13}Energy-corrected Ag partial phonon densities of states calculated from our RMC configurations. Successive curves have been raised by 2.5 units in each case to aid visibility. Dashed lines are guides to the eye that connect equivalent modes in each set of data.}
\end{figure}

One of the key results in our previous thermal expansion study was the decrease in Ag displacement frequencies with increasing temperatures, which provided a dynamical mechanism for overcoming the small changes in lattice enthalpies associated with thermal expansion of the lattice \cite{Goodwin_2008}. Here we can observe this softening directly by considering the Ag partial phonon density of states calculated for each temperature point in our RMC analysis; the corresponding plots are given in Fig.~\ref{fig13}. Again the form of these partial density of states contains an exponential background contribution due to ``noise'' in the RMC configurations. Not withstanding this effect, there is certainly an overall decrease in energy scale (the weighted mean frequencies corresponding precisely to those given in \cite{Goodwin_2008}). However, the lowest-frequency modes actually increase in energy with increasing temperature, despite the overall decrease in energy for Ag displacements. This indicates that there are modes with negative Gr{\"u}neisen parameters that involve Ag translations (remembering that the material as a whole shows PTE), and also that these modes occur at low energies. A natural candidate for these phonon modes is the collection of transverse vibrations of Co--CN--Ag--NC--Co linkages, which may also couple with RUMs of the [CoC$_6$] octahedra.

It is also interesting to note that the magnitude of the shifts in phonon frequencies can be moderately large: one can estimate from Fig.~\ref{fig13} that values of some mode Gr{\"u}neisen parameters $\gamma=-\partial(\ln \omega)/\partial(\ln V)$ are of the order of at least $\gamma=\pm3$. Consequently the magnitude of the overall thermal expansion behaviour ($\gamma$ of the order of 0.5, subject to the difficulty in obtaining a measure of the true compressibility \cite{Calleja_2008}) reflects a difference between competing thermal expansion effects of quite different types of lattice vibrations. This result contrasts sharply the behaviour of other NTE systems such as ZrW$_2$O$_8$ and Zn(CN)$_2$, where the overall behaviour is dominated by contributions from sets of phonon modes with similar Gr{\"u}neisen parameters \cite{Ravindran_2000,Zwanziger_2007}. This means that the extent of thermal expansion in Ag$_3$[Co(CN)$_6$] may be very sensitive to changes in composition, because subtle variations in the magnitudes of individual Gr{\"u}neisen parameters will have a much greater relative impact on their differences than on their sum.

We conclude this section by commenting that the similarity in structure of the Ag partial phonon density of states at each temperature point is encouraging in terms of the robustness of the PFTS approach in this case. Uncorrelated atomic motion produces a density of states that is an exponential decay function, with none of the reproducible structure seen in Fig.~\ref{fig13}. The level of detail of the phonon spectrum could certainly be improved if we were to extend our analysis to larger RMC configurations and/or larger configurational ensembles (while suffering the associated cost in additional computation time). However, the reproducibility we see at the level of analysis given here lends us some confidence that our results in terms of the key dynamical features and their general temperature dependencies are sufficiently robust for the purposes of this investigation.

\section{Conclusions}\label{conclusions}

This study has shown that the thermal expansion behaviour of Ag$_3$[Co(CN)$_6$] actually represents a somewhat complex synthesis of various different local effects. For example, it was always suspected that covalent bonding interactions would constrain thermal expansion along the Co--CN--Ag--NC--Co linkages, but we see here that actually the expansion of individual bonds is quite large, and that this PTE effect is counteracted by the effects of transverse vibrational motion (\emph{cf}.\ the local structure behaviour in Zn(CN)$_2$ \cite{Chapman_2005}). Similarly, the decrease in energy at higher temperatures seen for the Ag partial phonon density of states occurs as the (perhaps fortuitous) balance between vibrational modes with moderately-large negative \emph{and} positive Gr{\"u}neisen parameters.

Our analysis has also illustrated the complementary roles played by what one might term ``structural'' and ``dynamical'' flexibility of the Co--CN--Ag--NC--Co linkages. By ``structural flexibility'' what we mean is that the average geometry of these linkages is able to vary in order to preserve the [CoC$_6$] coordination geometries as the lattice parameters change with temperature. This was seen in particular in terms of thermal variation of the Ag--N--C bond angles. Alternatively, by ``dynamical flexibility'' we mean the ability to support low-energy transverse vibrational modes, which help constrain thermal expansion of the Co$\ldots$Co vector. We showed in particular the importance of RUMs, where transverse vibrational motion is correlated around individual metal centres in the form of polyhedral rotations and translations.

Our results strongly suggest that similarly atypical behaviour may be found in many other extended framework materials with flexible structures. Indeed in our parallel DFT study it emerged that the Au-containing analogue Ag$_3$[Co(CN)$_6$] is a very likely a colossal thermal expansion material \cite{Calleja_2008}. The same principles may yet prove relevant to other quite different materials, such as the increasingly diverse and widely-studied family of metal-organic frameworks (MOFs) \cite{Roswell_2004,Dubbeldam_2007}.

\ack{We acknowledge the University of Cambridge's CamGrid infrastructure for computational resources, Trinity College Cambridge for the provision of financial support to M.J.C. and A.L.G., and the E.P.S.R.C. (U.K.) for funding to J.S.O.E. and L.P. under ep/c538927/1.}


\begin{thebibliography}{10}
\bibitem{CN_peak_note}
Moreover, because the energy of the CN stretch is so much higher than thermal energy at all of the temperatures considered here, actually the phonon mode frequency calculated from the CN peak width is independent of temperature. We do not observe any significant change at all in the CN peak widths between 10 and 300\,K, and indeed the CN stretching frequencies consistently emerged in the region 35--40\,THz for each of our temperature points.
\end{thebibliography}
\end{document}